# Networks in Motion

Adilson E. Motter and Réka Albert

*Networks that govern communication, growth, herd behavior, and other key processes in nature and society are becoming increasingly amenable to modeling, forecast, and control.*

*Adilson Motter is a Professor of Physics and Astronomy at Northwestern University*

*Réka Albert is a Professor of Physics and Biology at Pennsylvania State University*

Twenty years ago Scott Feld, a sociologist from Stony Brook, wrote a scientific paper entitled "Why your friends have more friends than you do" [1]. Strange title for a paper, particularly because most people do not perceive themselves as being different from their friends. People often compare themselves with others to determine whether they are adequate. If people could count the number of friends their friends have, most of us would conclude that we do not have an adequate number of friends. Indeed, in any social group, it can be proven that the average number of friends of friends is larger than or equal to the average number of friends; moreover, the equality only holds for the very special case in which every person has the exact same number of friends.

Far from being a curiosity, this is intimately related to the percolation properties of networks. For example, if a contagious disease propagates through a social network and we are allowed to immunize a small fraction of the population, a naive approach would be based on selecting people at random and immunizing them. But this network idiosyncrasy suggests that selecting people at random and immunizing instead one of their acquaintances could lead to a far more effective approach [2]. A recent study, which followed the evolution of the H1N1 flu both in a group of randomly selected people and in a group of comparable size formed by friends named by those people, clearly indicated that the number of disease cases peaked about two weeks earlier in the group of friends (Physics Today, November 2010, p. 15). This illustrates that the same network property can also be used for the early detection of epidemic outbreaks. The question originally studied by Feld is in fact only one out of many broadly significant ones, ranging from self-organization and influence flow to systemic robustness, that can now be properly formalized within the emerging theory of complex networks.

Such questions are not unique to social networks and become more intriguing when they involve a reciprocal influence between the network structure and dynamical processes that take place on the network. For example, the island fox is a species unique to California's Channel Islands. About 150 years ago, feral pigs were introduced onto the islands, and, as for many other species in the absence of natural predators, their population grew fast. Even though pigs did not interact directly with foxes, the increase in the pig population was accompanied by a decrease in the population of foxes, bringing them close to extinction. This was caused by golden eagles, who, attracted by the abundance of pig preys, colonized the islands. The decline in fox population was then a consequence of the occasional interaction between eagles and foxes [3]. What is the solution to this problem? One may argue that it cannot simply be based on undoing what was done in the first place (i.e., on removing the pigs) because that would prompt the eagles to prey all the foxes. The golden eagle, on the other hand, is a protected species. This problem is challenging even though the relevant network consists of only 3 species—it could only be solved by the coordinated removal of two of the species. What to say then of much larger networks, such as those featured in Figure 1, which are often complex with respect to both their structure and dynamics, and which can exhibit yet other phenomena, including the propagation of cascading failures?

While problems like these may come from disparate fields, it is the theory of networks that unifies them. Physicists are at the forefront in developing concepts and theoretical approaches to address such problems. Common to many such problems is the fact that the network of interactions between the component parts can be just as important as the parts themselves in defining the properties of the system. The notion that the network matters has long been appreciated in physics—it is what makes the difference between fairly dissimilar allotropic materials, such as graphite and diamond, which are literally different networks of the exact same atoms. But many more possibilities arise in the study of networks of interacting organisms, computers, power stations, genes, financial institutions and so on, which depart strongly from the picture suggested by regular (or even disordered) local arrangements. These networks are both not constrained to only having local interactions and are heterogeneous, in particular with respect to the number of interactions per node (which is called *degree*, a generalization of the coordination number). In sparse random networks, for example, if the distribution of degrees for uniformly selected nodes is *P(k)*, the degree distribution for their network neighbors will be *kP(k)*, which, incidentally, is the basis of the sampling bias identified in networks of friends. But real networks, however complex, are not random. In fact, a substantial part of the recent interest in network research concerns the modeling of salient characteristics that make them different from random. In such systems the question is not whether but rather how the network organization influences systemic properties. The extent to which a complex network deviates from random provides insights into its organization, evolution, and function.

**The role of physicists**

The involvement of physicists in the area of networks was stimulated by the now decade-old discovery of ubiquitous structural properties in empirical networks, such as small-world and scale-free features, and the theoretical understanding of general mechanisms governing the emergence of these properties (Physics Today, November 2008, p. 33). In particular, numerous networks, including the Internet, human-contact networks underlying the transmission of diseases, air transportation networks, and metabolic networks, have been found to exhibit an approximately power-law distribution of degrees with a divergent variance (hence scale free). And it wasn't long until dynamical implications started to be identified. An early landmark was the demonstration that, for the simplest epidemic models, uncorrelated scale-free networks can exhibit a vanishing epidemic threshold in the limit of large population size [4], meaning that viruses or other agents with arbitrarily low spreading rates can persist in the population, thus exposing a previously unknown relation between network scaling and the vulnerability of human, animal, and computer networks. Accordingly, as the field of network science matures there is now increased added value to the disciplines it applies to.

Needless to say, the very study of networks is not new—food webs, social networks, and intracellular pathways have long been topics of research in the respective disciplines. The novelty is largely in the scale and complexity: the current science of networks is fueled by the availability of very large-scale information about these and other networks—world-wide Internet-based communities, whole-cell metabolic networks, global air transportation networks—which allows the application of methods from statistical physics (and also inspires theoretical studies, as illustrated in Box 1). An important trait of most such large networks is that they evolve and operate in a decentralized way, which in turn leads to the possibility of emerging dynamical phenomena, a recurrent topic in nonlinear physics.

**Box 1. The structure of a network** is often regarded as a collection of dots connected by lines, where the elements of the network are represented by the dots (nodes) and the interactions between them by the lines (edges)—cf. Figure 1. A fundamental question concerns the loss of interconnectivity upon the sequential removal of nodes or edges. In the control of an epidemic spread, for instance, the reason why targeted immunization is more efficient than random immunization is because the removal of highly connected nodes is more effective in fragmenting the network. Equally important is the reverse process: the emergence of large-scale connectivity—for example, through the addition of edges. In the simplest case (the Erdős–Rényi random-graph model), one starts with a large number $N$ of isolated nodes and adds one edge at a time between pairs of nodes randomly selected with uniform probability. A percolation transition then appears when, in the limit of large $N$, a finite fraction of the network becomes connected. This occurs when the ratio $r=m/N$ defined by the number of edges $m$ relative to the number of nodes $N$ reaches $r_c=1/2$. Slightly below this point, the sizes of the connected clusters follow a power-law distribution, with the largest being of order $\log N$. Slightly above, the size of the largest component is of order $N$ and grows linearly with $|r - r_c|$. Thus, this process shares properties with many continuous phase transitions studied in physical systems. But by changing the rules by which edges are added to the network, the transition point can be anticipated (i.e., $r_c<1/2$) or retarded (i.e., $r_c>1/2$). The latter, in particular, leads to interesting new transitions, dubbed "explosive percolation" because of their steep, seemingly discontinuous nature [5]. This case has been studied in detail for so-called Achlioptas processes, where a pair of edges is generated at each step but only the edge that minimizes the growth of the clusters is kept. The resulting delay causes the transitions to be sharper when they occur. Rigorous analyses have shown that these transitions are smooth, but they are described by nonanalytic scaling functions and the initial growth of the largest component scales with $|r - r_c|^\beta$, where $\beta << 1$ [5]. Explosive percolations are important for illustrating the impact of nonlocal information in the assignment of edges. This apparently academic exercise has implications for the study of spreading processes in networks, where the incidence of small events may be reduced but inadvertently at the price of increasing the impact of large ones.

**Self-sustained dynamics**

Collective behavior manifests itself in many forms in networks. Examples range from bursting neurons and phase-locked alternate current generators, to herd behavior in animal and social systems. A number of these processes can be idealized as manifestations of spontaneous synchronization, a phenomenon in which different elements of the network keep in pace with each other in a decentralized way. An important question concerns the properties of the network and its elements that can lead to such emergent behavior. Traditionally, this problem has been further simplified by considering models amenable to mathematical analysis, of which the best known example is the Kuramoto model—a minimal representation of a large population of coupled oscillators with nonidentical intrinsic frequencies, which was subsequently shown to be equivalent to certain arrays of Josephson junctions [6]. If each oscillator is coupled to all the others, under mild assumptions it can be shown that the system exhibits a critical coupling strength $K_c > 0$ above which a finite fraction of the network synchronizes. This phase transition is in many aspects similar to a percolation transition (Box 1)—in particular, as the coupling strength $K$ is further increased, an increased number of oscillators are recruited by the synchronous cluster. However, the problem becomes a lot more involved as soon as the network is allowed to be more general than all-to-all coupling. For example, the question of whether $K_c$ is finite or zero in scale-free networks remains unsettled, since mean-field calculations suggest it would vanish while numerical simulations are more consistent with it being strictly positive [7]. An exact solution, on the other hand, seems hard to come by with existing mathematical techniques. This illustrates the relative difficulty in addressing dynamical phase transitions in general networks of nonidentical nodes compared to their structural counterparts (Box 1).

In part because of this and other difficulties, much attention has been devoted to networks of identical or nearly identical dynamical units, particularly by means of the Pecora-Carroll model (which assumes that the nodes are diffusively coupled) [7]. Within this model, it is possible to show that heterogeneity in the distribution of connection strengths or in the dynamical units generally inhibits synchronization. However, it has been recently shown that different synchronization-inhibiting network properties, such as heterogeneous connectivity patterns and negative interactions, can compensate for each other when they coexist in the same network and in fact lead to a combined effect that facilitates synchronization [8]. This outcome is entirely due to collective behavior and cannot be anticipated from pairwise interactions between two nodes in the network. Related observations have implications for the number of nodes that can engage in a given activity. In particular, it can be shown that even when the network loses synchrony as the number of nodes is increased, it may regain synchrony when this number is further increased. Another fundamental result from these studies is that when the network structure is optimized for a given dynamical function, it often exhibits a cuspy landscape in the space of all possible networks (Figure 2), meaning that a small change in the network structure—such as a modification in the strength of an edge—can lead to a very different dynamical behavior [8]. This sensitive dependence of the collective dynamics on the structural parameters of the network can thus be interpreted as a network analog of chaos. This property is shared by other network processes and has implications for networks, such as biological ones, that have evolved under pressure to both optimize specific tasks and be robust to parameter perturbation.

**Spreading dynamics**

A different class of network processes that have received increased attention concerns spreading dynamics, which includes examples of innovation diffusion, consensus, emergence of norms, fads, riots, epidemic spreading, and propagation of failures. Many such processes can be conceptualized as either epidemic-type dynamics, where a node in an "infected" state may infect a neighbor independently of the status of the other nodes, or as cascading-type dynamics, where the change in the state of a node may depend on the reinforcement caused by changes in multiple other nodes. In both cases, the research focus is on the chain of events that takes place as the process unfolds and its relation to the properties of system.

The understanding of malware spread through the Internet, for instance, has been significantly advanced by the observation that this network exhibits a scale-free distribution of degrees at the autonomous system level and that this property is a determinant factor for the long-term persistence of computer viruses [4]. Analogous studies in human populations have remained elusive due to the significantly more challenging task of tracking human mobility and hence human contact networks. This challenge is in the process of being overcome, however, through at least three complementary approaches [9]. The first focuses on the global transportation of passengers, and has demonstrated the dominant role of air transportation in the spread of infectious diseases such as SARS. Other approaches to infer human contact patterns take advantage of cell phone use and currency bill tracking games, such as *Where's George* in the USA. These and other conceivable approaches that exploit recorded information left by people as they visit different locations are likely to lead to substantial new understanding about the role of social networks in the spread of infectious diseases. It may, in particular, yield understanding on how awareness of the presence of a disease changes people's behavior and possibly the outcome of an imminent outbreak.

The study of cascading processes, on the other hand, has provided valuable insights into what determines rare but large events. A key reference in this area has been the threshold model introduced by Watts [10], which assumes that the change in the state of a node from "normal" to a "modified" state—representing, e.g., a failure or change of opinion—will depend on the fraction of neighbors that have changed state. In its simplest form, the network is assumed to be random with a given degree distribution and the nodes in the network are assumed to have state transition thresholds (fraction of modified neighbors) determined by an independent distribution. This model can be studied rigorously in the thermodynamic limit (i.e., in the large population limit). Starting out with only one or few seed nodes in the modified state, it can be shown [10] that a finite fraction of the network will change to this modified state if (and only if)

$$\sum_k k(k-1)\rho_k P_k > z, \qquad (1)$$

where $P_k$ is the probability that a node has degree $k$, $\rho_k$ is the probability that a degree-$k$ node will change state if just one of its neighbors changes state (early followers), and $z$ is the average degree in the network (assumed to be finite). Note that the model accounts for the fact that different nodes may have a different number of connections and that the threshold for a node to change state will in general vary from node to node in the network. It follows from Eq. (1) that the vulnerability of the network to large cascades increases with increased heterogeneity of the thresholds and decreases with increased heterogeneity of the degrees (unless the seed nodes are purposely taken among the hubs). As a reference, it is useful to consider the hypothetical case in which all nodes are early followers (i.e., $\rho_k = 1$ for all $k$), so that Eq. (1) reduces to the condition for the network itself to have a macroscopic connected cluster. The conclusions may vary in cascades where other factors come into play, such as interdependencies [11], but one point is clear: the properties of the underlying network explain the empirical observation that a particular shock or innovation can trigger a large cascade while many apparently indistinguishable events lead to negligible consequences—it all depends on whether the initial event hits a percolating cluster of early followers.

**Dynamics of biological networks**

Living cells provide outstanding examples of collective behavior underpinned by networks of interactions. Cells use molecular-level biophysical networks to coordinate multiple functions, allowing them to respond to an ever-changing environment. A newly encountered signal, if it passes initial filtering mechanisms, may propagate through the network and result in a new cellular behavior or fate. Even in the absence of external signals, the cross-regulation among gene products may result in the possibility of multiple steady-state and/or dynamic equilibria — also known as *attractors*, these are orbits towards which trajectories converge over time and as such represent the long-term behavior of the system. The emergent dynamics of intra-cellular regulatory networks have been on physicists' minds since the late 1960s, when Stuart Kauffman introduced random Boolean networks as simple prototypical models of network-based collective behavior [12]. Although the initial models invoked a mixture of homogeneity and randomness, and some of their specific results did not hold the test of time, they paint a compelling picture: biological networks are poised near the boundary between the two extremes of frozen and chaotic dynamics. In this critical region, the state change of one node propagates on average to one other node. This conclusion is preserved if more realistic assumptions about the structure of the underlying network, or about the logic of the regulatory relationships, are made (Physics Today, March 2009, p. 8).

Genome-wide experimental methods now identify interactions among thousands of components. However, progress toward understanding the collective behavior emerging from these networks has been slow. The reason is that quantitative dynamical characterization of every reaction participating even in a relatively simple function of a single cell requires a concerted and decades-long effort. In this context, Boolean models have emerged as serious contenders for striking a balance between economy and realism when modeling real biological processes, from cell cycle to embryonic development, cell differentiation, and programmed cell death (Figure 3). In multiple instances the predictions of these models, such as those concerning the effects of node disruptions, were validated experimentally [13]. Recent work on Boolean dynamic models focuses on the potential effects of stochasticity and uncertainty on the predicted attractors. For example, introducing multiple time scales in the model, to capture the observed variability in the duration of the processes represented as edges in the interaction network, reduces the total number of attractors but does not affect the steady-state attractors. The vast majority of cellular behaviors are indeed steady states or trajectories toward a steady state, suggesting that robustness to timing variability is a general characteristic of biological systems that the models should also exhibit.

The increasing availability of network-based dynamic models also allows probing the connection between network structure and dynamics. Another pioneer of Boolean models of gene regulatory networks, Rene Thomas, conjectured that the necessary condition for sustained oscillations in genetic networks is the existence of a negative feedback loop, and that the necessary condition for multiple steady states is the existence of a positive feedback loop [14]. These conjectures have been proven correct in several dynamic frameworks, both continuous and discrete. Intra-cellular regulatory networks do exhibit both positive and negative feedback loops, thus the potential for both dynamic behaviors exists. But which feedback loops are actually active at any given instance? Answering this question will necessitate the incorporation of the regulatory logic into the model of the network structure.

**Control of perturbed networks**

Halting an epidemic spread or the loss of biodiversity are both examples of network control problems. Networks have evolved or have been designed to operate stably, but as these and other examples illustrate, this balance can be lost when they are perturbed. Most previous studies of postperturbation control have focused on epidemiological problems, which continue to be as important as they have ever been. Recent network studies indicate, however, that postperturbation control can also be developed for network failures as diverse as those underlying financial crises, power outages, and genetic diseases. The concept is that, because most of the large-scale damage is determined by the cumulative effect of the spread of failures rather than by any isolated event, these failures may be largely controllable even after the perturbation that causes them—a bankruptcy, malfunction, or mutation—has already hit the system.

One promising approach consists in exploiting the multi-stability inherent to a large number of complex networks [15]. For example, perturbed food webs can undergo extinction cascades, but the targeted suppression of specific species can drive the network to a new stable state in which most or all species survive. Similarly, living cells that are unable to reproduce due to faulty metabolic pathways can in principle be rescued by rebalancing the metabolic fluxes through the targeted suppression of specific genes. Moreover, for certain types of cancer, genetic interventions can make the disease state disappear, leaving the healthy state as the only possibility. Even though these examples concern natural networks, it is expected that research on the control of network failures will have significant impact on the development of self-healing technological networks, such as "smart" power-grid networks. While there is a big leap between theoretical and practical applications, the possibility of recovering from failures in real time may also have implications for the control of financial crises.

**Outlook**

Much of contemporaneous physics is based on the paradigm that macroscopic phenomena do not depend on the microscopic details of the process. They do depend, however, on the underlying network of interactions. Because of the widespread effects of networked technologies, people are now predisposed to engaging with the importance of network concepts in their lives. So are physicists.

The network enterprise concerns, on the one hand, the identification and analysis of network features that are relevant to the particular phenomenon of interest. But, with so many conceivable possibilities, what if we simply fail to look for the right features? It just so happens that researchers have been thinking about this too. In fact, in the case of purely structural features, two exploratory methods have been devised [16], which can identify patterns not anticipated by pre-conceptions. One based on maximum likelihood techniques and the other on an integrative visual analytics approach, they both aim at resolving the internal structure of complex networks by organizing the nodes into groups that share something in common, even if we do not know a priori what that thing is. At first this may sound a little like the Deep Thought's "answer to the ultimate question of life, the universe, and everything" in Douglas Adams' fiction comedy series, except that in this case we can actually identify the question itself—a simple example of which is given in Figure 4. This is of course only the very tip of the iceberg. A broader undertaking concerns the development of similar exploratory approaches that can also systematically account for dynamical behavior, which, truth be said, remains widely unexplored.

On the other hand—beyond the need to reduce complexity and understand the workings of *existing* networks—one ought to exploit networks to *build* things with desirable properties that may not be readily available. Examples include new areas such as synthetic biology and microfluidics, which could be revolutionized by rational circuitry design, but also very classical areas such as traffic and materials research. In the latter case, an important precedent is the recent progress in the study of metamaterials, which are engineered materials that gain their properties from their structure (hence network) rather than their composition. These materials often exhibit properties, such as negative refractive index and acoustic shielding, not found in any conventional material. Yet, they have so far been limited to rather conservative network architectures, and this is only in part for the excellent reason of easier fabrication. One can only speculate the possibilities that arbitrary networks of polypeptides, nucleic acids or carbon nanotubes could lead to. As fabrication techniques improve, and allow advances such as the customized assembly of bio-molecules, the limiting factor is really one's ability to anticipate the interplay between the underlying network and the resulting systemic properties, which now lurk undiscovered.

Taken together, the quest for how *network* quantitative differences lead to qualitative differences is an overarching theme of current research. In retrospect, this epitomizes much of Philip W. Anderson's now-classic essay on the hierarchical structure of science and the argument for why "more is different" [17], but now from the vantage point of network theory.

[1] S. L. Feld, *Am. J. Sociol.* **96**, 1464 (1991).

[2] R. Cohen, S. Havlin, D. ben-Avraham, *Phys. Rev. Lett.* **91**, 247901 (2003).

[3] F. Courchamp, R. Woodroffe, G. Roemer, *Science* **302**, 1532 (2003).

[4] R. Pastor-Satorras, A. Vespignani, *Phys. Rev. Lett.* **86**, 3200 (2001).

[5] D. Achlioptas, R.M. D'Souza, J. Spencer, *Science* **323**, 1453 (2009); R. A. da Costa, S. N. Dorogovtsev, A. V. Goltsev, J. F. F. Mendes, *Phys. Rev. Lett.* **105**, 255701 (2010);

P. Grassberger, C. Christensen, G. Bizhani, S.-W. Son, M. Paczuski, *Phys. Rev. Lett.* **106**, 225701 (2011); O. Riordan, L. Warnke, *Science* **333**, 322 (2011).

[6] S. H. Strogatz, *Physica D* **143**, 1 (2000).

[7] S. N. Dorogovtsev, A. V. Goltsev, J. F. F. Mendes, *Rev. Mod. Phys.* **80**, 1275 (2008).

[8] T. Nishikawa, A.E. Motter, *Proc. Natl. Acad. Sci. USA* **107**, 10342 (2010).

[9] V. Colizza, A. Barrat, M. Barthélemy, A. Vespignani, *Proc. Natl. Acad. Sci. USA* **103**, 2015 (2006); M. C. González, C. A. Hidalgo, A.-L. Barabási, *Nature* **453**, 779 (2008); D. Brockmann, L. Hufnagel, T. Geisel, *Nature* **439**, 462 (2006).

[10] D. J. Watts, *Proc. Natl. Acad. Sci. U.S.A.* **99**, 5766 (2002).

[11] S. V. Buldyrev, R. Parshani, G. Paul, H. E. Stanley, S. Havlin, *Nature* **464**, 1025 (2010).

[12] S. A. Kauffman, *The Origins of Order: Self-Organization and Selection in Evolution*, Oxford U. Press, New York, NY (1993).

[13] R. Zhang, M. V. Shah, J. Yang, S. B. Nyland, X. Liu, J. K. Yun, R. Albert, T. P. Loughran, *Proc. Natl. Acad. Sci. U.S.A.* **105**, 16308 (2008); J. Saez-Rodriguez et al., *PLoS Comput. Biol.* **3**(8), e163 (2007); O. Sahin et al., *BMC Syst. Biol.* **3**, 1 (2009).

[14] R. Thomas, R. d'Ari, *Biological Feedback*, CRC Press, Boca Raton, FL (1990).

[15] S. Sahasrabudhe, A. E. Motter, *Nat. Comm.* **2**, 170 (2011); A. E. Motter, N. Gulbahce, E. Almaas, A.-L. Barabási, *Mol. Syst. Biol.* **4**, 168 (2008), A. Saadatpour, R.-S. Wang, A. Liao, X. Liu, T. P. Loughran, I. Albert, R. Albert, *PLoS Comput. Biol.* **7**(11): e1002267 (2011).

[16] M. E. J. Newman, E. A. Leicht, Proc. Natl. Acad. Sci. USA 104, 9564 (2007); T. Nishikawa, A. E. Motter, *Sci. Rep.* **1**, 151 (2011).

[17] P. W. Anderson, *Science* **177**, 393 (1972).

[18] For further reading, see: R. Cohen, S. Havlin, *Complex Networks: Structure, Robustness and Function* (Cambridge University Press, Cambridge, UK, 2010); A. Barrat, M. Barthélemy, A. Vespignani, *Dynamical Processes on Complex Networks* (Cambridge University Press, Cambridge, UK, 2008).

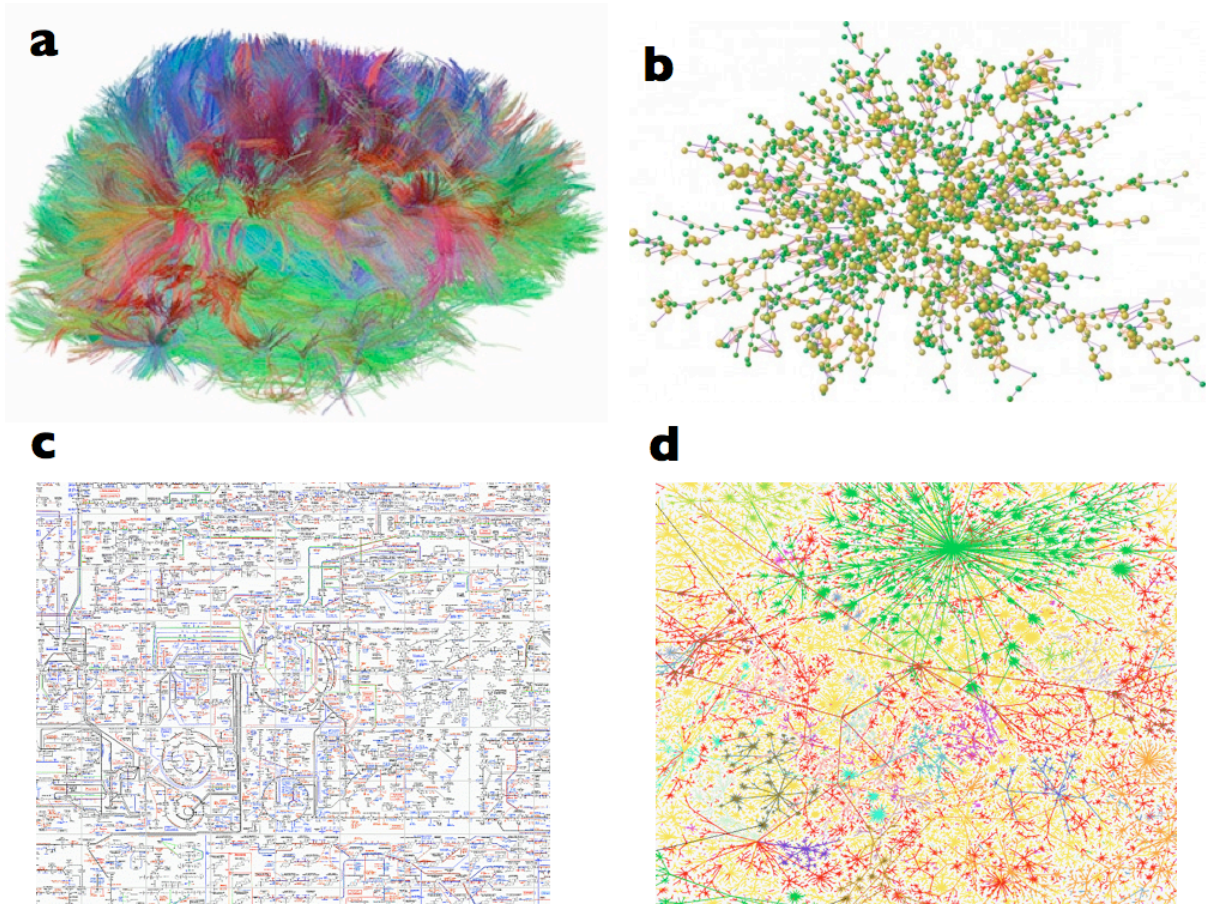

**Figure 1. Examples of networks** of current scientific interest. (a) Brain wiring of a human adult resolved using diffusion spectrum imaging, where each line represents a bundle of nerve fibers (Credits: V. J. Wedeen/Harvard). (b) Social network used to establish the social influence of body weight, where yellow nodes correspond to obese people (Credits: N.A. Christakis and J.H. Fowler, NEJM 35, 370 (2007)). (c) Part of the network of known metabolic pathways in living cells (Credits: Boehringer Mannheim/Biochemica). (d) Part of the network of major Internet service providers colored by IP addresses. (Credits: B. Cheswick and H. Burch/Lucent Technologies).

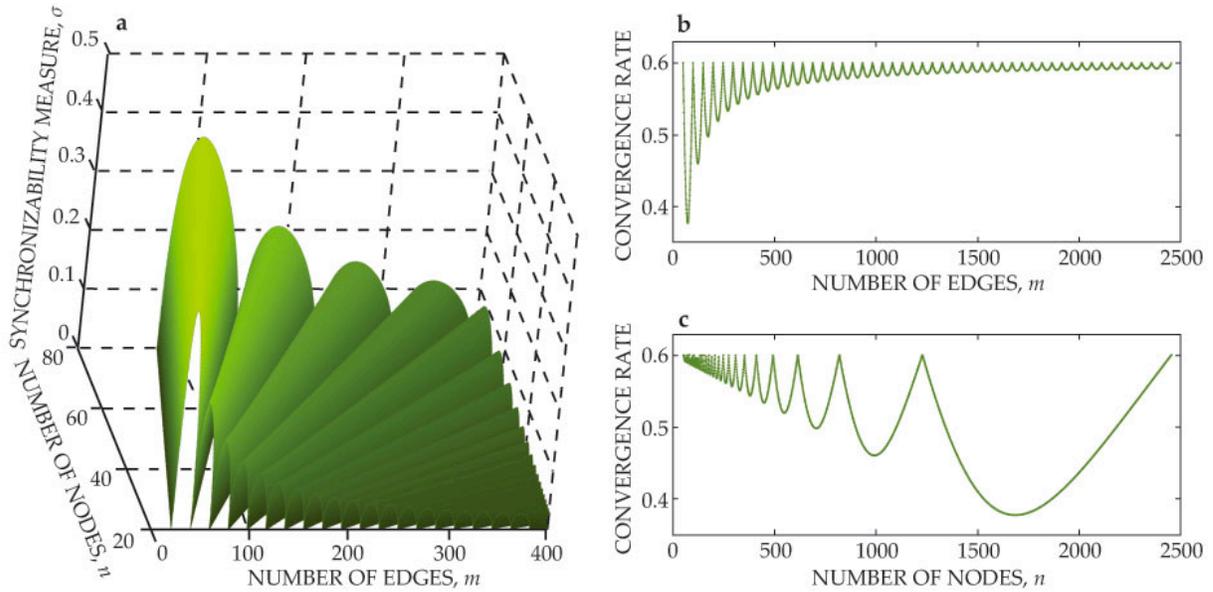

**Figure 2. Landscape defined by networks** that optimize synchronizability. (a) The figure shows the standard deviation $\sigma$ of the eigenvalues of the network's coupling matrix, which is a measure of the propensity of the network to synchronize, for the networks of $n$ nodes and $m$ edges that minimize $\sigma$. A higher stability of synchronous states corresponds to a smaller $\sigma$, indicating that collective behavior is facilitated along the cusps, where the derivative of $\sigma$ with respect to $m$ and $n$ both diverge. This has direct implications for physical quantities, such as convergence rates. (b, c) Corresponding rate of convergence to a synchronous state as a function of the number of edges for networks of 50 nodes and (b) as function of the number of nodes for networks of 50x49 edges (c). Similar, highly non-monotonic behavior is also observed for other physical quantities (Adapted from T. Nishikawa et al., PNAS 107, 10342 (2010); B. Ravoori et al., PRL 107, 034102 (2011)).

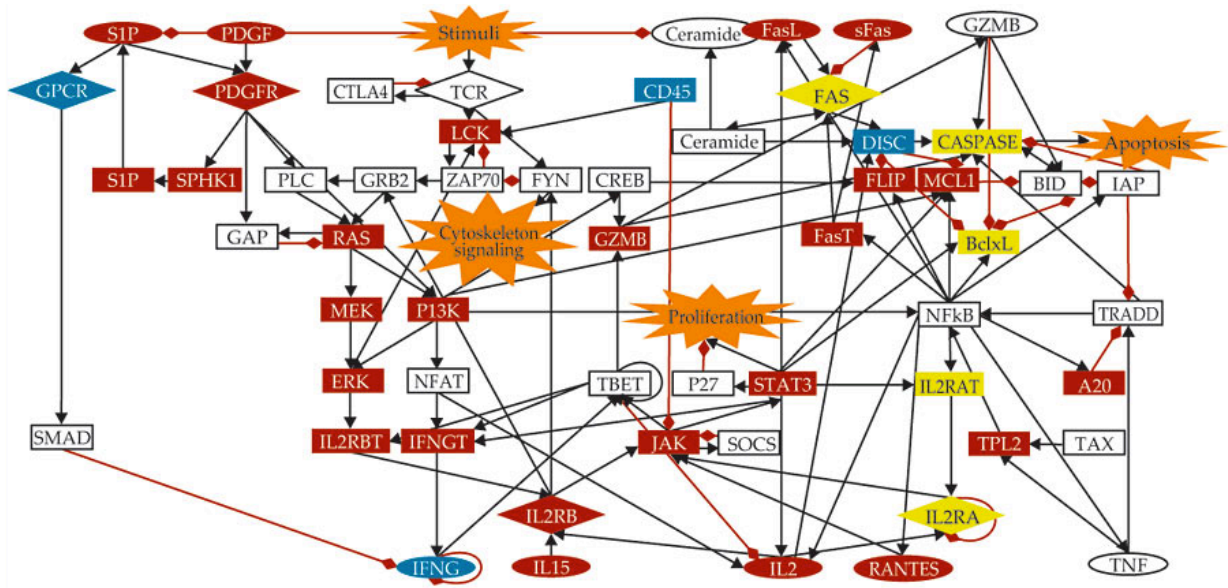

**Figure 3. Signaling network of T cells** responsible for their programmed cell death (apoptosis) in response to certain stimuli in their environment. The nodes in the network represent proteins and the directed edges indicate interactions between the corresponding proteins. T cells participate in immune responses and depend on this signaling network to control their population. Malfunction of this network leads to a form of blood cancer called large granular lymphocytic leukemia. Boolean modeling indicates that targeting certain proteins, such as IL15, PDGF and NFkB, can control the abnormal survival of T cells that causes the disease. (Adapted from R. Zhang et al., PNAS 105, 16206 (2008)).

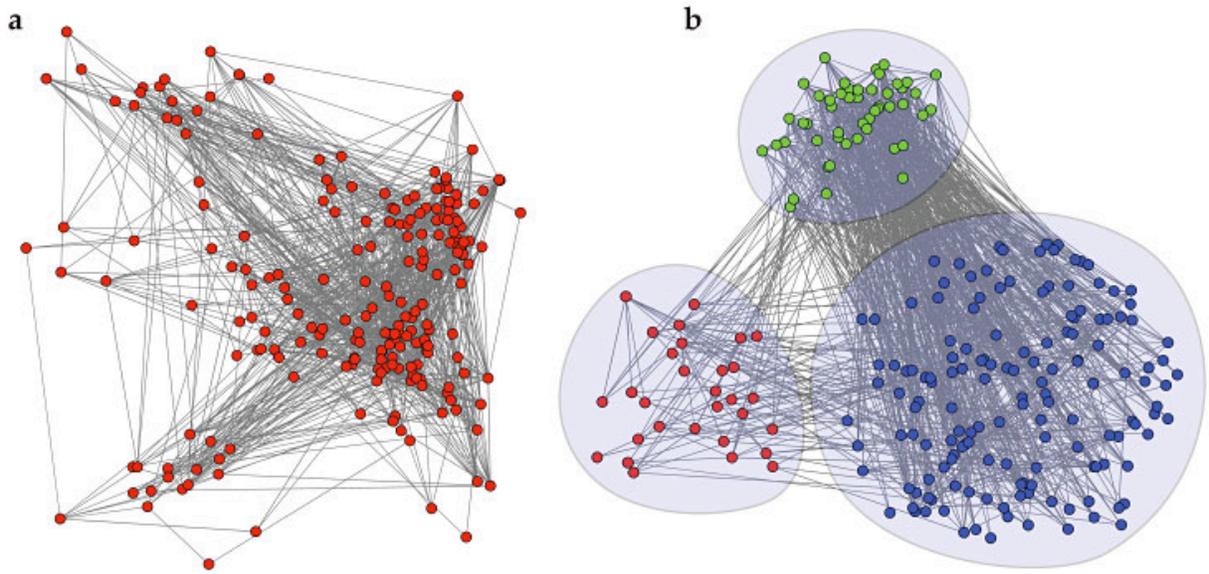

**Figure 4. Resolving internal network organization** to reveal hidden structural patterns. (a) Given the network on the left, the exploratory method identifies three groups of nodes that share common structural properties without specific *a priori* information about what defines such groups. Starting from a high dimensional "node property space" that summarizes all relevant properties of the nodes in the network, the method is based on integrative analysis of multiple two-dimensional projections from that space, which can be inspected by eye, to then separate the nodes into groups clustered around common properties. (b) *A posteriori* inspection of the groups, laid out on the right, reveals that they are defined by lower-degree nodes with lower-degree neighbors (red), lower-degree nodes with higher-degree neighbors (blue), and higher-degree nodes (green), and hence cannot be resolved by a single node property. (Credits: T. Nishikawa et al. Sci. Rep. 1, 151 (2011)).